# Comparability of Driving Automation Crash Databases

Noah Goodall

Virginia Transportation Research Council, USA, noah.goodall@vdot.virginia.gov

June 26, 2024

**Abstract**

*Introduction:* This paper reviewed current driving automation (DA) and baseline human-driven crash databases and evaluated their comparability. *Method:* Five sources of DA crash data and three sources of human-driven crash data were reviewed for consistency of inclusion criteria, scope of coverage, and potential sources of bias. Alternative methods to determine vehicle automation capability using vehicle identification number (VIN) from state-maintained crash records were also explored. *Conclusions:* Evaluated data sets used incompatible or nonstandard minimum crash severity thresholds, complicating crash rate comparisons. The most widely-used standard was "police-reportable crash," which itself has different reporting thresholds among jurisdictions. Although low- and no-damage crashes occur at greater frequencies and have more statistical power, they were not consistently reported for automated vehicles. Crash data collection can be improved through collection of driving automation exposure data, widespread collection of crash data form electronic data recorders, and standardization of crash definitions. *Practical Applications:* Researchers and DA developers may use this analysis to conduct more thorough and accurate evaluations of driving automation crash rates. Lawmakers and regulators may use these findings as evidence to enhance data collection efforts, both internally and via new rules regarding electronic data recorders.

**Disclaimer:** The views and opinions expressed in the article are those of the author and do not necessarily reflect the official policies or positions of any agency of the Commonwealth of Virginia.

# 1 Introduction

Many vehicles on the road are equipped with automated driving features. These features are often classified based on their alignment with SAE Levels of Automation. The National Highway Traffic Safety Administration (NHTSA) classifies Levels 1–2 systems as Advanced Driver Safety Systems (ADAS), and Levels 3–5 as Automated Driving Systems (ADS) (National Highway Traffic Safety Administration, 2023d). The term "driving automation" here refers to vehicles equipped with Levels 2–5 automated driving features.

Although manufacturers have accumulated millions of miles of ADS testing (California Department of Motor Vehicles, 2023b) and billions of miles of Level 2 ADAS operation (Karpathy, 2020), there have been few studies evaluating the on-road crash risk of these systems. To evaluate crash risk, researchers must have records of driving automation (DA) crashes, exposure (often expressed as DA miles traveled or hours in motion), and comparable crash rates for conventional vehicles. Because of limited DA operation, crash sample sizes are small, and comparisons to high-quality conventional vehicle fatal crash databases are impossible for many years at current DA testing rates (Kalra & Paddock, 2016). Researchers instead must compare DA crash rates to lower-quality low- and no-damage conventional vehicle crash data.

There are several databases of DA and non-DA crashes. With few exceptions, these data sources have incompatible or vague crash inclusion standards, making direct comparison between datasets difficult. For example, an estimated 60% of property-damage-only crashes do not appear in police crash databases (Blincoe et al., 2015). This can lead to misleading results when comparing minor crashes required by the California Department of Motor Vehicles (DMV) Autonomous Vehicles Testing Program (California Department of Motor Vehicles, 2018), many of which are not reported to police, with police-reported crashes in the NHTSA Crash Report Sampling System (National Highway Traffic Safety Administration, 2023a). DA exposure data is also extremely limited, especially for Level 2 ADAS. As crash severity can range from no-damage to fatal, careful comparison between databases regarding low-severity crashes is essential for accurate analysis.

When data is available, it can often be biased due to some DA technologies used exclusively on freeways with lower baseline crash rates or in urban areas with higher baseline crash rates. Other technologies, especially ADS, have been tested only at low-speeds in urban environments free of rain and snow (Kusano et al., 2023). Other DA software updates may be rolled out to the safest drivers first, further skewing results.

Researchers, DA developers, and regulators continue to assess DA safety, often by comparing DA and conventional vehicle crash rates. These data are stored in different databases with different standards, which may not allow direct comparison. The purpose of this study is to provide context around the characteristics of these databases, with particular emphasis on the direct comparability of different databases. This study contributes to the literature by providing guidance to researchers, regulators, and practitioners who may be integrating these databases in an attempt to evaluate DA safety, or who may be evaluating the quality of other DA safety assessments. When applicable, guidance for how to filter or normalize data for direct comparison is provided. A discussion of alternative and non-public databases is provided, as researchers may seek or obtain access to these sources. Finally, recommendations to improve existing databases are presented.

## 2    Literature Review

The scientific literature on driving automation crash databases can be categorized into exploratory studies and benchmark development, and driving automation safety comparisons against human-driven vehicles.

### 2.1    Exploratory Analyses and Benchmark Development

Researchers have explored driving automation crash databases and identified trends and preliminary findings without attempting to compare crash rates with human-driven vehicles. These include exploratory analyses (Das et al., 2020; Quintero, 2022), text mining (Alambeigi et al., 2020; Boggs et al., 2020), crash rate calculation (Leilabadi & Schmidt, 2019), and crash severity modeling (Wang & Li, 2019).

Other researchers have attempted to establish crash rate benchmarks for driving automation. Waymo evaluated existing sources for human-driven crash rates based on police reports, with adjustments for road type, region, reporting thresholds, and underreporting (Scanlon et al., 2023). This represents one of the first papers to identify the problem of vehicle-level crash rates when comparing driving automation and human-driven datasets. Driving automation datasets record crashes and mileage at the individual vehicle level, whereas most human-driven datasets record mileage for all vehicles but crashes only once regardless of how many vehicles were involved. When crashes and mileages are estimated for all vehicles, not just a sample, then crashes must be counted as vehicle-crashes, where a two-vehicle collision is counted twice. Waymo's paper focused exclusively on average crash rate benchmarks, was not peer-reviewed, and is susceptible to a conflict of interest as these benchmarks were later used to assess Waymo's own safety (Kusano et al., 2023).

Another study on driving automation crash rate benchmarks considered alternatives to average human-driven crash rate. Using data from a naturalistic driving study, Goodall (2021b) calculated the crash rates for model driving, defined as "sober, rested, attentive, and cautious," to be 33% lower than for all driving, and proposed this as a potential adjustment factor to conventional benchmarks. The study also considered crash rate per unit time instead of the traditional mileage, and incident/injury/fatality rates of existing professionally-driven or computer-driven transport modes such as buses, passenger rail, and elevators.

### 2.2    Crash Rate Comparisons

Other studies have attempted to assess DA crash rates by comparing them with baseline crash rates using various driving automation and conventional vehicle crash databases and exposure metrics. Schoettle and Sivak (2015) conducted the first study comparing ADS crash rates to conventional vehicle crash rates. At the time, only 11 ADS crashes over 1.2 million miles had been reported. The authors found higher crash rates for ADSs, but the results were not statistically significant. The authors compared the mostly low- and no-damage crashes in the California DMV database with police-reported crashes in NHTSA's General Estimates System (GES), adjusted based on NHTSA's estimate of a 60% unreported crash rate. Both Favarò et al. (2017) and Banerjee et al. (2018) conducted studies using similar methods, but compared low-impact ADS crashes with GES police-reported crashes without adjusting for unreported crashes.

Dixit et al. (2016) compared ADS crashes in the California DMV data between September 2014 and November 2015 with California Highway Patrol conventional vehicle crash records. The authors noted that California Highway Patrol's statewide crash rate covered only State, U.S., and Interstate roads, and not the local streets on which ADSs were tested at the time.

The authors failed to adjust for crash severity, as California DMV records include all crashes while California Highway Patrol records only include crashes that met reporting criteria of $1000 minimum damage (Blincoe et al., 2023), were reported to police, and entered into their database.

In a study funded by Google, Blanco et al. (2016) compared Google's ADS crash rates with data from the Second Strategic Highway Research Program Naturalistic Driving Study (SHRP 2 NDS) (Dingus et al., 2016). Using a non-public dataset from Google, they were able to identify Google crashes that met police-reporting thresholds. The authors defined these thresholds as crashes resulting in more than $1500 in damage, significant impacts (e.g. delta-v > 20 mi/hr or acceleration > 1.3g excluding curb strikes), or contact with a large animal or sign. These were compared to identically filtered crash records in SHRP 2 NDS. They also compared all Google crashes with all crashes in the age-weighted SHRP 2 NDS data, including those below police reporting thresholds.

Teoh and Kidd (2017) expanded this analysis to include Google ADS crashes and mileage nationally, using monthly activity reports published on Google's website (Davies, 2015). These monthly reports were discontinued and removed from Google's website in late 2016 in conjunction with the spinoff of their ADS group as Waymo (Kovach, 2017). Google crash rates were filtered for crashes that met police reporting thresholds regardless of actual reporting, and these were compared against conventional vehicle crash reports. They found several Google crashes that met thresholds yet were not reported to police, suggesting that the comparison between reported and reportable crashes may introduce bias. The authors further compared Google crash rates with SHRP 2 NDS for all crashes regardless of severity. The authors, however, appear to have used unweighted SHRP 2 NDS data. This introduced overcounting, as SHRP 2 NDS oversampled teen and elderly drivers who are known to have greater crash risk (Antin et al., 2019).

Goodall (2021a) compared ADS and conventional vehicle crash rates for struck-from-behind crashes. The datasets compared were the California DMV ADS crashes and age-weighted SHRP 2 NDS across all severity levels. Crash rates were further filtered by driving environment in the SHRP 2 NDS data, as most of the ADSs conducted operations in urban (Cruise) or business/industrial (Waymo) settings.

Waymo published a brief white paper comparing insurance claim rates of Waymo ADS vehicles to rates for human-driven vehicles (Di Lillo et al., 2023). The study investigated claims incurred by Waymo as a proxy for fault. Exposure data for huma-driven vehicles was not measured directly but rather estimated FHWA estimates. It is difficult to evaluate the accuracy of the findings due to the lack of peer review and proprietary underlying data.

In another paper, Waymo compared the crash rate of its driverless operation (which they refer to as "rider-only") against human-driven vehicles (Kusano et al., 2023) using human-driven crash rate benchmarks of insurance claims data (Di Lillo et al., 2023) and crash records (Scanlon et al., 2023) developed in earlier reports. In much of the analysis, the authors attempted to compensate for underreporting of crashes in the human-driven estimates. The authors attempted to control for the driverless operational design domain, in which vehicles were limited to "non-limited access roads with speed limits up to 50 mph and parking lots without restrictions on maneuvers" while avoiding "severe weather conditions, such as thick fog, heavy rain, or blowing sand" (Kusano et al., 2023, p. 4).

In a methods article, Goodall (2024) used Tesla's self-reported Autopilot ADAS system and non-Autopilot crash rates to demonstrate techniques to deal with sparse data in crash data analysis. Tesla's reported crash rates while running Autopilot were compared to crash rates of Teslas without Autopilot after controlling for higher rates of freeway use by vehicles with Autopilot. Because Tesla uses non-standard crash definitions, a comparison to NHTSA Crash Report Sampling System (CRSS) data was impossible, but a comparison to airbag crash rates in the SHRP 2 NDS was attempted.

A summary of DA to conventional vehicle crash comparisons is shown in Table 1. As demonstrated in the table, many studies have relied on data that incompatible, and produced misleading results. Some of these analyses could have been improved using existing data and more careful normalization and filtering.

**Table 1 Driving Automation Crash Studies and Comparisons in the Literature**

| DA Metric | DA Crashes per Million Miles (95% CI) | Baseline Metric | Baseline Crashes per Million Miles (95% CI) | Source |
|---|---|---|---|---|
| All crashes | 9.1 (4.5,16.3) | Police-reportable crash estimates | 4.1 (3.5, 4.7) | (Schoettle & Sivak, 2015) |
| All crashes | 23.8 | Reported crashes | 2.0 | (Favarò et al., 2017) |
| All crashes | 30.5–8,843 | Reported crashes | 2.0 | (Banerjee et al., 2018) |
| All crashes | 21.2 | Reported by CHP, limited coverage | 0.5 | (Dixit et al., 2016) |
| Waymo police-reportable in California | 2.19 (0.44, 6.39) | Police-reported in Mountain View | 6.06 (5.93, 6.18) | (Teoh & Kidd, 2017) |
| Google police-reportable | 4.57 (2.09, 8.68) | Police-reported | 3.59 (2.31, 4.87) | (Teoh & Kidd, 2017) |
| *All Waymo crashes | 9.7 (5.8, 15.1) | SHRP 2 all crashes | 27.6 (25.8, 29.5) | (Teoh & Kidd, 2017) |
| *All Waymo rear-end struck | 7.1 (3.9, 11.9) | SHRP 2 rear-end struck | 2.7 (2.1, 3.3) | (Teoh & Kidd, 2017) |
| All Waymo crashes | 8.8 (2.6, 22.8) | All SHRP 2 | 26.8 (23.9, 30.1) | (Blanco et al., 2016) |
| *All Waymo crashes | 8.8 (2.6, 22.8) | All SHRP 2, age-adjusted | 20.2 (17.7, 23.0) | (Blanco et al., 2016) |
| Waymo police-reportable | 3.2 (0.4, 11.4) | SHRP 2 police reported | 1.4 (0.9, 2) | (Blanco et al., 2016) |
| Waymo police-reportable | 3.2 (0.4, 11.4) | SHRP 2 police reported, age-adjusted | 0.9 (0.5, 1.5) | (Blanco et al., 2016) |
| Waymo police-reportable | 3.2 (0.4, 11.4) | SHRP 2 police reportable | 8.2 (6.9, 9.7) | (Blanco et al., 2016) |
| *Waymo police-reportable | 3.2 (0.4, 11.4) | SHRP 2 police reportable, age-adjusted | 5.8 (4.7, 7.0) | (Blanco et al., 2016) |
| Waymo police-reportable | 3.2 (0.4, 11.4) | NHTSA police reportable | 4.2 (2.8, 9.9) | (Blanco et al., 2016) |
| *Tesla Autopilot, airbag or active restraint system | 0.45 | Tesla non-Autopilot, airbag or active restraint system | 0.49 | (Goodall, 2024) |
| *Rear-end struck | 17.2 (14.2, 20.7) | SHRP 2 rear-end struck | 3.6 (3.0, 4.3) | (Goodall, 2021a) |
| *Waymo supervised property damage insurance claims | 0.17 (0.06, 0.37) | Swiss Re property damage insurance claims | 3.17 (3.16, 3.18) | (Di Lillo et al., 2023) |
| *Waymo supervised injury insurance claims | 0.09 (0.02, 0.25) | Swiss Re injury insurance claims | 1.09 (1.08, 1.09) | (Di Lillo et al., 2023) |
| *Waymo driverless property damage insurance claims | 0.78 (0.16, 2.27) | Swiss Re property damage insurance claims | 3.26 (3.24, 3.27) | (Di Lillo et al., 2023) |
| *Waymo driverless injury insurance claims | 0.00 (0.00, 0.95) | Swiss Re injury insurance claims | 1.11 (1.10, 1.12) | (Di Lillo et al., 2023) |
| *Waymo driverless injury | 2.1 | Police-reported locally | 4.85 | (Kusano et al., 2023) |
| *Waymo police-reported | 0.4 | Police-reported, adjusted for underreporting | 2.78 | (Kusano et al., 2023) |

* Metrics use similar crash definitions and are directly comparable.

## 3 Driving Automation System Database Characteristics

There are several DA crash databases. This section provides an overview of the most widely used databases, including examples of their usage in the literature and a summary of their comparability challenges and unique characteristics.

### 3.1 NHTSA Standing General Order Level 2 ADAS Crash Data

In June 2021, NHTSA issued a Standing General Order (SGO) requiring several vehicle manufacturers and operators to report certain crashes on publicly accessible roads involving vehicles using Level 2 ADAS and Levels 3–5 ADS (National Highway Traffic Safety Administration, 2021b). The order was amended in August 2021 (National Highway Traffic Safety Administration, 2021a) and again in April 2023 (National Highway Traffic Safety Administration, 2023d). This section discusses the records of Level 2 ADAS crashes, whose reporting requirements differ from Levels 3–5 ADS crashes.

Level 2 ADAS involves the simultaneous vehicular control of "lateral and longitudinal vehicle motion control subtasks" with a driver responsible for monitoring "object and event detection and response" (SAE International, 2021). An example of a Level 2 ADAS is simultaneous lane-centering and adaptive cruise control functions. Vehicles with Level 2 ADAS have been sold to the public since at least 2015 (Dikmen & Burns, 2016).

The main requirements to report a Level 2 ADAS crash is that the Level 2 ADAS was active at any point 30 seconds prior to the crash, and that "the crash results in any individual being transported to a hospital for medical treatment, a fatality, a vehicle tow-away, or an air bag deployment or involves a vulnerable road user" (National Highway Traffic Safety Administration, 2021b). Therefore, many low-severity crashes or non-injury crashes are not required to be reported.

Most manufacturers do not have a mechanism which automatically alerts them to vehicle crashes. This is evident in the crash records themselves, as Tesla, for example, receives the majority of their reports from telematics data while Honda receives the majority from consumer complaints (Tucker, 2022).

When crashes are reported, it is not always certain when Level 2 ADAS systems are engaged. Approximately 10% of Level 2 ADAS crashes in the SGO database had the field "Automation System Engaged?" coded as "Unknown, see Narrative" with the narrative redacted.

Although the SGO requires crashes beginning on July 9, 2021 (ten days after the June 29 date of the original SGO), crashes were reported from as far back as 2019. This suggests that manufacturers may interpret reporting requirements differently and may report some crashes that do not meet crash severity thresholds out of an abundance of caution.

The SGO has no requirement for reporting system exposure, either in terms of miles traveled or hours in motion under L2 ADAS. Without exposure data, and without completeness of reporting, the SGO L2 ADAS data is mostly useful in investigating month over month trends and has limited utility in generating crash rates or comparing to other crash databases.

### 3.2 NHTSA Standing General Order ADS Crash Data

NHTSA issued a Standing General Order (SGO) in June 2021 for operators to report ADS vehicle crashes occurring on public roads (National Highway Traffic Safety Administration, 2021b). See the prior section for an overview of the SGO. This section discusses how the SGO addresses vehicles with ADS, defined by NHTSA as vehicles operating in automation levels 3–5.

ADS is distinguished from Level 2 ADAS by the systems' responsibility for object and event detection and response, with the human driver required as fallback (SAE International, 2021). Vehicles with ADS are not available for purchase according to NHTSA (2023d), but have been tested on public roads in the United States continuously since 2010 (Beiker, 2014).

Unlike manufacturers of Level 2 ADAS, ADS operators were not given a crash severity threshold, and instead were required to report all crashes occurring on public roads. NHTSA defines crashes in two ways. First, operators are required to report instances where the vehicle contributed to or was alleged to contribute to a crash that "results or allegedly results in any property damage, injury, or fatality" (National Highway Traffic Safety Administration, 2021b). NHTSA goes on:

> "For clarity, a subject vehicle is involved in a crash if it physically impacts another road user or if it contributes or is alleged to contribute (by steering, braking, acceleration, or other operational performance) to another vehicle's physical impact with another road user or property involved in that crash."
> (National Highway Traffic Safety Administration, 2021b)

It is possible to have a crash where a physical impact is involved but no damage is incurred, e.g., bumping a cyclists rear tire as occurred in Report 30413-4170 in the NHTSA SGO ADS Crash Data (National Highway Traffic Safety Administration, 2023e). It is unclear from NHTSA's definition of "crash" whether this type of incident would need to be reported. Of the 283 unique ADS crash records, 27 (8.2%) were flagged as no injuries and no property damage. From analysis of the narrative field, some of these crashes may have incurred damage and were possibly miscoded, others involved hit and runs with minor damage to the ADS vehicle, others involved minor damage such as fender indentations, while others involved no-injury no-damage interactions with vulnerable road users (VRUs). NHTSA defines a VRU as anyone who is not an occupant of a motor vehicle with more than three wheels (National Highway Traffic Safety Administration, 2021b).

Based on NHTSA's broad definition of crash in the ADS SGO, there appear to be reported crashes that do not involve property damage or injury. These crashes would not be captured in some other databases such as the NHTSA CRSS property-damage-only crash estimates.

NHTSA (2022) clearly states that, in their interpretation, vehicles with ADS are not available for consumer purchase. By default, then, all ADS must be operated by developers, and all applicable crashes should be knowable and reportable by the developer.

### 3.3    California DMV Autonomous Vehicles Program

The California DMV initiated an autonomous vehicle testing program in 2014, requiring all ADS developers testing on public roads in California to obtain a Manufacturer's Testing Permit (California Department of Motor Vehicles, 2018). Permit holders were required to fill out and submit the OL 316 form with details of any crash in California involving a vehicle operating under the permit, regardless of whether it was in autonomous mode or not. Permits are required for developers testing Levels 3–5 ADS, but are not required for Level 2 ADAS (California Department of Motor Vehicles, 2018).

California's system is unique in three ways. First, they require crash records for all crashes involving permitted vehicles, regardless of whether they were operating under autonomous control at the time of the crash. Section 5 of the OL 316 form (California Department of Motor Vehicles, 2021) includes a data field for whether the vehicle was in autonomous or conventional mode at the time of the crash. No information is provided for precisely when a vehicle was most recently in autonomous mode, although several studies have been able to classify crashes as occurring shortly after transitioning to conventional control based on the crash narratives (Blanco et al., 2016; Goodall, 2021a; Schoettle & Sivak, 2015). In instances where a vehicle transitioned from autonomous control to manual control while stopped, and then was struck from behind while stopped, these studies classified these crashes as occurring in autonomous mode crashes regardless of the selection in the OL 316.

The second way California's system is unique is that it requires permit holders to report their autonomous-mode miles traveled annually (California Department of Motor Vehicles, 2023a). Mileages are listed by month, by permit holder company, and by individual vehicle. This is one of the only sources of ADS exposure data and it is widely used to estimate ADS crash rates (Banerjee et al., 2018; Dixit et al., 2016; Favarò et al., 2017; Goodall, 2021a; Schoettle & Sivak, 2015).

Finally, California is the only jurisdiction to require reports of system disengagements. A disengagement of the ADS refers to a deactivation of autonomous mode when a system failure is detected, or when safe operation requires disengagement initiated either by the test driver or, in the case of a driverless vehicle, the system itself (California Department of Motor Vehicles, 2018). The relationship between disengagements and crashes is mixed (Khattak et al., 2021). Attempts to assess ADS safety from disengagement rate could create a disincentive for test drivers to initiate disengagements, even in potentially risky situations. Safety analyses from disengagement rates should use caution when drawing conclusions from disengagement rates.

Recent year crash, mileage, and disengagement data are available on California DMV's website (California Department of Motor Vehicles, 2023a, 2023c), while later years are available upon request. California DMV data is archived by the Autonomous Vehicle Operation Incident Dataset project (Zheng et al., 2023).

## 4    Conventional Vehicle Database Characteristics

Crash rates of driving automation systems are often compared to baseline metrics of human-driven vehicles, referred to here as conventional vehicles (Blanco et al., 2016; Goodall, 2021a; Schoettle & Sivak, 2015; Teoh & Kidd, 2017). This section reviews the main conventional vehicle crash databases and their limitations.

### 4.1    *NHTSA Crash Report Sampling System and Fatal Accident Reporting System*

NHTSA releases annual estimates of police-reported crashes. Between the 1970s and 2015, property damage and injury crashes were based on the National Automotive Sampling System General Estimates System (NASS GES). In 2016, the GES was replaced by the Crash Report Sampling System (CRSS). Fatal crashes were recorded in the Fatal Accident Reporting System (FARS) throughout this period. This analysis focuses on the CRSS-FARS combination, as CRSS was in use during the majority of driving automation systems testing. Both CRSS and FARS can be queried using the Fatality and Injury Reporting System Tool (FIRST) (National Highway Traffic Safety Administration, 2023f).

While FARS contains over 140 data fields of what are assumed to be all fatal crashes in the United States, the CRSS instead relies on a sampling system to estimate the number of police-reported crashes, and relies solely on police reports (National Highway Traffic Safety Administration, 2023b).

Not all crashes are reported to the police. To estimate the number of unreported crashes, NHTSA first used a telephone survey asking respondents about their reported and unreported crashes over the prior 12 months (M. Davis and Company, Inc., 2015). NHTSA estimated that 15.4% of injury crashes and 31.9% of property-damage-only crashes were never reported to police (M. Davis and Company, Inc., 2015). A second NHTSA study (Blincoe et al., 2023) of 2019 data updated these figures by considering also the crashes reported to the police but never officially filed. These non-filings may have been instances where the police were called but were unable to respond, crashes not meeting state reporting thresholds (state minimum damage varies between $0 and $3000 (Blincoe et al., 2023)), or other reasons. This study estimated that 31.9% of injury crashes and 59.7% of property-damage-only crashes were never filed (Blincoe et al., 2015). Across all crashes, the unreported rate was 53.2%. A third estimate can be obtained from the SHRP 2 NDS in which passive data collection instruments were installed in 3,542 drivers' vehicles over a period of 1–2 years (Dingus et al., 2016). From this dataset, Blanco et al. (2016) estimated that 84% of property-damage-only crashes were unreported. Their estimates relied on self-reporting from SHRP 2 NDS participants and may not have captured crashes that were reported by the other party. As a result, the researchers considered 84% to be an upper estimate of the actual unreported crash rate, with low estimates of 35% and mid-estimates of 60% from telephone surveys (M. Davis and Company, Inc., 2015) and an economic impact report (Blincoe et al., 2015), respectively. A fourth estimate from analysis of 6.7 million hours of traffic camera video in Virginia found that 55% of observed crashes did not appear in state crash databases (Bareiss, 2023).

Analysis that compares NHTSA FARS/CRSS crash rates to databases with different inclusion standards should adjust the crash rates so that both are comparing either police-reported crash rate or total crash rate estimates. For example, the California Autonomous Vehicles Testing Program requires companies to report low-speed crashes, even those without any associated property damages (California Department of Motor Vehicles, 2018). When comparing with NHTSA police-reported crash rates, only DA crashes reported to police should be compared.

### 4.2    SHRP 2 Naturalistic Driving Study

Between 2012 and 2014, the Second Strategic Highway Research Program Naturalistic Driving Study (SHRP 2 NDS) collected over 2 petabytes of video, kinematic, and audio data from 3,542 drivers using passive data collection systems installed in participants' personal vehicles (Antin et al., 2019; SHRP 2 NDS, 2020). This remains the largest naturalistic driving dataset with over 33 million miles driven and 36,000 crash, near crash, and baseline events.

Crash and near crash data were collected during short intervals prior to and following critical events. Critical events were triggered by kinematic or crash sensors surpassing acceleration thresholds, or by drivers' pressing a button in the vehicle. Trained analysts reviewed data to classify and code events.

SHRP 2 NDS is one of the only sources of unreported and no-damage crashes. Crashes are divided into four severity levels, from Level 4 tire strikes to Level 1 airbag deployments, towing, and VRU injuries (Virginia Tech Transportation Institute, 2015). SHRP 2 NDS also records periodic baseline events where cameras record for a brief period not in response to a crash or near-crash event. These baseline events allow researchers to monitor driver behavior and road conditions in non-event scenarios. The baseline events provide a better understanding of conventional vehicle exposure data, as time of day, environment (e.g., urban/rural), and road classification. This allows researchers to estimate crash rates in different environments that better correspond to often highly specific DA usage.

As part of their emphasis on high-risk drivers, the SHRP 2 NDS intentionally oversampled both young and elderly drivers. Any attempt to estimate national crash rates from SHRP 2 NDS must re-weight the data to reflect the United States driving population. Resampling weights from Blanco et al. (2016) are reproduced in Table 2.

**Table 2 Age Group Sample Weights for SHRP 2 Naturalistic Driving Study Data**

| Age | Weight | Percentage in SHRP 2 NDS | Percentage of US Licensed Drivers | Million miles driven | Weighted million miles driven |
|---|---|---|---|---|---|
| 16-24 | 0.32 | 37 | 12 | 12.9 | 4.1 |
| 25-39 | 1.53 | 17 | 26 | 6.4 | 9.8 |
| 40-54 | 2.33 | 12 | 28 | 4.6 | 10.7 |
| 55-74 | 1.35 | 20 | 27 | 6.3 | 8.5 |
| 75+ | 0.5 | 14 | 7 | 3.4 | 1.7 |
| Totals | - | 100 | 100 | 33.6 | 34.8 |

Table 3 summarizes the coverage of the automated and conventional vehicle crash databases.

**Table 3 Database Coverage by Crash Severity.**

| Source | No damage | Damage, not police reportable | Police-reportable | Police-reported | Injury | Fatal |
|---|---|---|---|---|---|---|
| SGO Level 2 ADAS | VRU-involved | If known and VRU, tow-away, or airbag involved | | | If known | |
| SGO ADS | | | ✓ | ✓ | ✓ | ✓ |
| California DMV | | ✓ | ✓ | ✓ | ✓ | ✓ |
| NHTSA CRSS/FARS | | | | ✓ | ✓ | ✓ |
| SHRP 2 NDS | ✓ | ✓ | ✓ | ✓ | ✓ | ✓ |

# 5 Alternative Data Sources

In addition to the publicly available databases described in prior sections, there are alternative data sources that may provide additional context.

## 5.1 Manufacturer-supplied VIN Records

The Insurance Institute for Highway Safety (IIHS) has in the past conducted safety research using publicly available state crash records (Cicchino, 2017, 2018). To determine whether a given vehicle was equipped with certain ADAS features, some vehicle manufacturers voluntarily provided to the Highway Loss Data Institute "special samples" (Insurance Institute for Highway Safety, 2017) of vehicle identification numbers (VINs) of those vehicles equipped with these optional features. These VINs could be cross referenced with state crash records to determine whether a crashed vehicle was equipped with an ADAS feature. In a comment to proposed federal legislation, IIHS acknowledged that this type of research would be impossible without manufacturer cooperation.

NHTSA has made efforts to obtain vehicle-level data on ADAS installations directly from manufacturers. In 2018, NHTSA and several automakers established the Partnership of Research in Traffic Safety (PARTS) to facilitate the exchange of data on ADAS equipped rates based on VINs of crash-involved vehicles (Kolly & Czapp, 2019). An independent third party, currently the MITRE Corporation, operates the group and ensures data privacy. As of December 2021, automakers have reported ADAS equipment for 47 million model-year 2015–2020 vehicles (Partnership for Analytics Research in Traffic Safety, 2021). Auto industry participation is voluntary and decisions regarding the scope and objectives of research projects require consensus of the entire Governance Board, with each automaker assigned a single vote (Partnership for Analytics Research in Traffic Safety, n.d.). Although PARTS has published favorable research results in the past (Partnership for Analytics Research in Traffic Safety, 2022), it is unclear whether members under this structure can refuse to allow publication of research findings that could be perceived as unfavorable. Regardless, raw data is not made available to the public, severely limiting the research potential of this valuable data.

For state and federal databases that include the VINs of vehicles involved in crashes, some information on automation technology may be obtained from NHTSA's Product Information Catalog and Vehicle Listing (vPIC) database (National Highway Traffic Safety Administration, 2023c). For certain manufacturers and model years, NHTSA searches for automation technology that was available as a standard or optional feature (National Center for Statistics and Analysis, 2023). While data is not available on an individual vehicle like the IIHS database, vPIC is publicly available. VINs can be searched in bulk via an API by downloading vPIC as a standalone database. Gajera et al. (2023) used this approach to analyze the prevalence of partial automation technologies in fatal crashes. VINs are not available in NHTSA CRSS but are included in many state crash databases as well as NHTSA FARS for fatal crashes.

## 5.2   Tesla Autopilot

Tesla introduced the combined functionality of adaptive cruise control and lane centering (called Autosteer) as a feature on Model S vehicles in October 2015 (Barry, 2021). The system is referred to as Autopilot and is now an available option on all Tesla vehicles with the necessary hardware.

Unlike many vehicle manufacturers, Tesla collects vehicle crash data in near real-time using cellular network telematics (Tucker, 2022). Tesla has released reports four times per year stating rates of miles per crash for vehicles with and without Autopilot active (Tesla, Inc., 2023c). These reports cover the period beginning July 2018 through December 2022, as of this writing in July 2023. (Between July 2018 and March 2021, non-Autopilot crashes were split into two separate categories. These figures were used in the literature (Goodall, 2024), but have since been reclassified on Tesla's website.) Tesla updated their historical crash rates in January 2023 to correct faulty data "where no airbag or other active restraint deployed, single events that were counted more than once, and reports of invalid or duplicated mileage records" (Tesla, Inc., 2023c).

Tesla includes only crashes involving deployment of an airbag or "other active safety restraint" in the Tesla vehicle (Tesla, Inc., 2023c). Tesla provides additional context: "In practice, this correlates to nearly any crash at about 12 mph (20 kph) or above, depending on the crash forces generated" (Tesla, Inc., 2023c).

This crash definition is difficult to compare across databases. While the SHRP 2 NDS records airbag deployments, they do not record active restraint systems. Furthermore, different manufacturers may have different thresholds for active restraint system activation. The 12 mi/hr threshold is also difficult to apply without guidance regarding correlation to maximum point-of-force deceleration or delta-v. A collision with a fixed barrier, for example, generates greater forces on a vehicle than a collision with a stopped vehicle due to the stopped vehicle's ability to absorb the crash forces (Evans, 1994). It is also not clear if 12 mi/hr refers to the speed of the subject vehicle alone or the speed differential between two vehicles.

Although Tesla uses the same crash definitions for both Autopilot and non-Autopilot crash rates, they are exposed to dissimilar road classifications. Tesla Autopilot was intended for use only on controlled access freeways (Tesla, Inc., 2023a), and freeway crash rates per mile are 30% of the non-freeway crash rate according to a naturalistic driving study (Goodall, 2024). Any comparison of Tesla Autopilot crash rates would need to control for Autopilot's greater (but not exclusive) freeway use and its associated lower crash rate. For an example of an analysis of Autopilot crash rate using normalized data, see Goodall (2024).

### 5.3 Tesla Full-Self Driving

In October 2020, Tesla released a software update enabling a "Full Self-Driving Beta" (FSD) feature for limited vehicles (Bonifacic, 2020). The update expanded the lane keeping and adaptive cruise control features in Autopilot, notably adding the ability to navigate onto freeway offramps and respond to traffic signals (Monticello & Barry, 2021). Most notably, this expanded the use of Tesla driving automation systems from predominantly freeway use with Autopilot, to more urban surface street uses.

In their 2022 Impact Report, Tesla cited statistics that vehicles with FSD engaged and active crashed at a rate of 0.31 crashes per million miles traveled, compared to 0.18 for Autopilot and 0.68 for Tesla vehicles with no active safety features engaged and running (Tesla, Inc., 2023b).

The FSD crash rate may be biased based on the method in which the technology was released. Tesla assesses the safety profiles of Tesla drivers using a metric called a Safety Score (Tesla, Inc., 2021). Drivers are rated on a 0–100 scale based on factors such as forward collision warnings, hard braking, aggressive turning, and unsafe following. The exact metrics and score calculations are available on Tesla's website (Tesla, Inc., 2021).

FSD software updates were first released to drivers with safety scores of 100 on October 2020 (Stone, 2023). By June 2022, 21 months later, driver with scores as low as 91 reported receiving invitations to activate FSD (Stone, 2023). By September 2022, the minimum required score was set to 80 (Ali, 2022), and on November 24, 2022, the Safety Score minimum was removed (Nedelea, 2022). Unsurprisingly, drivers' Safety Scores appeared to be correlated with higher crash rates. According to a unit-less bar chart in Tesla's 2022 Impact Report, drivers with a safety score between 0–60 crashed 3.1 times as often as drivers with a score between 91–100 on a per mile basis (Goodall, 2023). The crash rate may have applied to FSD operation as well, given evidence that drivers using FSD must intercede in frequent, inconsistent intervals (Nordhoff et al., 2023).

Putting the 2022 FSD crash rate into context, for half the year, only the safest drivers were permitted to use FSD. For 11 months of 2022, FSD drivers were known by Tesla to have considerably lower crash rates than drivers prohibited from activating FSD features. This may have produced a considerable bias in the form of lower FSD crash rates. The precise effect is impossible to calculate without further data, as the distribution of drivers' Safety Scores was not released.

The threshold for FSD crashes is not mentioned in Tesla's Impact Report (Tesla, Inc., 2023b), but readers are referred their safety reports (Tesla, Inc., 2023c). While the safety reports do not address FSD crashes, the crash threshold are defined there as deployment of an airbag or other active safety restraint, roughly correlated to crashes at speeds of 12 mi/hr or greater, depending on forces involved. The FSD crashes likely have the same reporting thresholds as those for Tesla Autopilot crashes.

The FSD crash rate may be further affected by recalls (Shepardson, 2024), necessitating careful consideration of FSD version and data collection times when analyzing crash rates.

## 6 Strategies to Improve DA Crash Databases

Evaluating DA safety from crash rates remains a challenging problem. In the absence of thorough national data, consistent crash definitions, and detailed exposure data, researchers have been limited to using crash surrogates, modeling, and non-comparable crash rates. This section presents strategies to improve the state of DA crash data collection.

### 6.1 Exposure Data

Exposure data, expressible in either miles traveled or hours in motion with automation engaged (Goodall, 2021b), is necessary to estimate crash rates. While there are several sources of exposure data for conventional vehicles, exposure data for DA systems is currently limited to California DMV records for vehicles with ADS. Mileage by vehicle or by manufacturer would greatly increase the value of NHTSA SGO data, providing nationalized crash rates for vehicles with ADS. Mileages (or hours of operation) should be classified by automation operating status, e.g., autonomous mode on or off, as well as road classification and geography, e.g., urban vs. rural. This data could be easily collected, as manufacturers are already required to report all ADS crashes regardless of severity. Members of Congress have called for NHTSA to collect mileage as part of the SGO since 2023 (Mullin et al., 2024; Mullin & Pelosi, 2023), suggesting that mileage can be collected under NHTSA's regulatory authority. Level 2 ADAS exposure data is more difficult to collect at a national level, and would provide less value as only a small proportion of Level 2 ADAS crashes are currently collected.

### 6.2 Event Data Recorders

Most new vehicles are voluntarily equipped with event data recorders (EDRs). For vehicles so equipped, they are required to record certain data at specified intervals in the seconds immediately prior to an activation event (Federal Register, 2006). Many crashes in the NHTSA SGO databases have associated EDR data submitted to NHTSA, but these data are not made public. The EDR data includes two fields directly relevant to crash analysis: lateral and longitudinal delta-v and airbag deployment. These could be used to determine an objective measure of crash severity, allowing direct comparison of crashes of similar severity across databases without relying on subjective measures such as police reported and police reportable.

New rules could also require EDRs to capture information on vehicle automation technologies. Most Level 2 ADAS vehicles involve the integration of separate ADAS functions such as lane keeping, lane centering, adaptive cruise control, and collision avoidance. Were EDRs to record not only the operational status of these ADAS functions but also their behavior prior to the crash (e.g., adaptive cruise control desired speed), researchers could conduct much more thorough analysis of DA vs. conventional vehicle crash rates. With enough data, researchers could even compare crash rates of identical vehicle years and models both with and without these systems active, eliminating most sources of noise. Expanding EDR requirements would probably require new rulemaking by NHTSA, the process of which has historically taken on average 2.6 years for new rules (Public Citizen, 2016, p. 35).This timeline varies depending on complexity of the rule, its economic impact, and perceived political climate (Potter, 2017), with averages ranging between 6 months and 9.1 years depending on agency and rule type (Public Citizen, 2016, p. 34).NHTSA could instead coordinate voluntary agreement among automakers; this was the approach taken when introducing automatic emergency braking, which NHTSA estimated was three years faster than the rulemaking process (National Highway Traffic Safety Administration, 2016).

EDR data could also be collected for conventional vehicle crashes. Under the FAST Act of 2015, Congress permits the downloading of EDR data for research purposes provided any personally identifiable information and VIN are not disclosed (Davis, 2015). States could require crash investigations to download and record EDR data as part of police investigations into all crashes. Because manufacturers have nonstandard and often complex means to download EDR data, NHTSA could require manufacturers to install simple, uniform methods of data retrieval. A USB port installed under the dash of all new vehicles to seamlessly download EDR data is just one example.

*6.3 Crash Definition Standardization*

Automated driving represents a relatively small portion of all driving, and safety studies are forced to analyze low-severity crashes in order to have adequate sample sizes. Unfortunately, low speed crashes suffer from inconsistent data collection, with different standards for police-reporting across states and vague distinctions when estimating the value of property damage. An objective metric for crash severity would help researchers more precisely compare low- and no-damage crashes that make up the bulk of DA crashes. One possible metric is maximum delta-v, defined as the maximum value of the cumulative change in velocity from moment of impact to 0.3 seconds after, as required in existing EDRs (Federal Register, 2006). Delta-v has been used a crash severity metric for decades (Robinette et al., 1994), and has shown to be highly correlated to injury level and probability of fatality.

Regardless of the selected metric, a formal standardization effort could be led by SAE or ISO. Similarly, NHTSA and states could informally standardize crash severity by reporting estimated including such metrics in conventional vehicle and DA crash databases.

# 7   Conclusions

Despite manufacturers having accumulated millions of miles of ADS testing and billions of miles of Level 2 ADAS operation, safety assessments of DA systems have been confined to limited, often non-compatible datasets. Due to small sample sizes of DA crashes, analyses have relied on low- and no-damage crashes which suffer from inconsistent reporting and few reliable metrics of severity. These challenges have made it difficult for researchers to directly compare the on-road crash rates of DA systems with those of conventional vehicles.

This study evaluated existing public datasets for both DA and conventional vehicle crash rates. The main challenges identified were inconsistent and vague crash standards for low-severity DA and conventional crashes, and a severe lack of exposure data for DA crashes.

Three strategies to improve data collection were presented: the collection of DA exposure data in miles traveled or hours in motion at a national level, release of EDR data and enhancement of EDR collection to include driving automation system statuses, and standardization of crash severity metrics to include maximum delta-v.

Researchers conducting DA safety analyses may use these findings to design more sophisticated studies, normalizing data in various database to allow direct, unbiased comparisons. Regulators may use these findings to conduct their own safety studies and better evaluate studies in the literature. Lawmakers and regulators may use the findings as evidence to enhance data collection efforts, both internally and via new EDR rules. DA developers may use these studies in internal and external safety studies to more accurate assess the safety of their driving automation systems.

## Funding
This work was sponsored by the Virginia Department of Transportation. The views and opinions expressed in this article are those of the author and do not necessarily reflect the official policies or positions of any agency of the Commonwealth of Virginia.

## CRediT Authorship Contribution Statement
**Noah Goodall:** Conceptualization, Methodology, Formal analysis, Investigation, Writing – original draft, Writing – review & editing, Visualization.

## Declaration of Competing Interests
The author served as a paid expert witness for the plaintiffs in litigation filed against Tesla.